\documentclass[prl,twocolumn,showpacs,superscriptaddress]{revtex4-1}
\usepackage{graphicx}
\usepackage{amsmath,amssymb,braket}
\usepackage{hyperref}
\usepackage{color}

\newcommand{\hidden}[1]{}

\begin{document}
\preprint{aniso/H{\"u}bner}

\title{Electron g-Factor Anisotropy in Symmetric $\mathbf{(110)}$-oriented GaAs Quantum Wells}

\author{J. H{\"u}bner}
\email[email: ]{jhuebner@nano.uni-hannover.de}
\author{S. Kunz}
\author{S. Oertel}
\affiliation{Institute for Solid State Physics, Leibniz
Universit{\"a}t Hannover, Appelstr. 2, 30167 Hannover, Germany}
\author{D. Schuh}
\affiliation{\mbox{Institute for Experimental and Applied Physics,
Universit{\"a}t Regensburg, 93040 Regensburg, Germany}}
\author{M. Pochwa{\l}a}
\author{H. T. Duc}
\author{J. F{\"o}rstner}
\author{T. Meier}
\affiliation{\mbox{Department of Physics and CeOPP, Universit{\"a}t Paderborn, Warburger Str. 100, 33098 Paderborn, Germany}}
\author{M. Oestreich}
\affiliation{Institute for Solid State Physics, Leibniz
Universit{\"a}t Hannover, Appelstr. 2, 30167 Hannover, Germany}

\date{\today}

\begin{abstract}

We demonstrate by spin quantum beat spectroscopy that in undoped symmetric (110)-oriented GaAs/AlGaAs single quantum wells even a symmetric spatial envelope wavefunction gives rise to an asymmetric in-plane electron Land{\'e}-g-factor. The anisotropy is neither a direct consequence of the asymmetric in-plane Dresselhaus splitting nor of the asymmetric Zeeman splitting of the hole bands but is a pure higher order effect that exists as well for diamond type lattices. The measurements for various well widths are very well described within $14\times 14$ band $k\cdot p$ theory and illustrate that the electron spin is an excellent meter variable to map out the internal -otherwise hidden- symmetries in two dimensional systems. Fourth order perturbation theory yields an analytical expression for the strength of the g-factor anisotropy, providing a qualitative understanding of the observed effects.
\end{abstract}

\pacs{78.55.Cr,78.47.jd,78.20.Ci,71.18.+y}

\maketitle


Symmetry is a fundamental principle which runs through all fields of sciences like a common thread. The balance of proportions is attracting  great interest ever since reaching from Euclid's geometry theorems and the Archimedes lever principle in ancient times to Mandelbrot sets in present day mathematics and parity violation in modern particle physics. At the beginning of the last century the topic was significantly pushed by Emmy Noether's discovery of the deep connection between symmetry and conservation laws \cite{noether1918} and the classification of nearly all entities in today's physics in terms of its symmetry properties is a very powerful and widely applied method in a vast number of fields. Among the plethora of interesting physical observables the pure quantum mechanical entity \textit{spin} in connection with the relativistic effect of spin-orbit interaction (SOI) \cite{thomasn1926} bears an exceeding connection to symmetry. In a free atom, SOI can break the degeneracy of states with the same orbital wave function owing opposite spins. In solids, however, such a splitting interferes with crystal symmetry. The most prominent example is the conduction band Dresselhaus splitting in zinc-blende (ZB) type lattice semiconductors \cite{dresselhauspr1955}, which is not present in their diamond lattice type equivalents \cite{golubprb2004}. The alteration of the symmetry allows a clear assignment of the investigated spin properties to the symmetry at hand and the change of  symmetry properties on micro- and macroscopic scales is easy to produce in solid state physics by the introduction of low dimensional structures, potential gradients, or the choice of peculiar crystallographic quantization axes. This fact has boosted a great interest in recent semiconductor spintronic research \cite{zuticrmp2004, fabianaps2007, wupr2010} since crystal symmetry yields a control on the spin dynamics \cite{bychkovjpc1984, ganichevprl2004, tarasenkoprb2009, eldridgeprb2011, ducprb2010} and contrariwise the entity \textit{spin} yields jointly with the time-reversal breaking property of a magnetic a unique meter variable to probe internal symmetries which might be inaccessible by other means. 


In this letter, we exploit the intriguing property that quantum wells (QW) grown with their quantization axis along the low symmetry $[110]$ direction belong to the same symmetry class $C_{2v} $ as asymmetric $(001)$-oriented QWs. However, the spatial part of the wavefunction remains symmetric in growth direction for the $(110)$-oriented structure and it is only the spin dependent part, i.e., the Dresselhaus and Zeeman contributions which senses the symmetry reduction. The introduction of a two dimensional confinement changes  for $(001)$-oriented bulk GaAs crystals the primary ZB symmetry from $T_{d}$ to $D_{2d} $. This gives rise to the anisotropy of in- and out-of-plane g-factors \cite{ivchenko1992, jeunesst1997, malinowskiprb2000, pfefferprb2006}. Further suppression of symmetry operations -- leaving only the identity operation, a two-fold rotation axis and two mirror-planes -- yields $C_{2v} $ symmetry. However, the arrangement of the mirror-planes can be achieved in two different ways for ZB-based QWs: a) by a gradient along a (001)-quantization direction which constrains all mirror planes to contain the quantization axis or b) by the choice of the $[110]$ axis as growth and quantization direction which places one mirror plane in the middle of the QW. The astonishing fact is, even though being clear  if the point group operations are transferred to the g-factor tensor, that in case a) the electron spin acquires an additional dynamic due the asymmetric envelope wavefunction in conjunction with SOI \cite{kalevichjetp1992, eldridgeprb2011}, whereas in case b) the envelope wavefunction is fully symmetric for the electrons at the conduction band minimum. Nevertheless, the spin still acquires an additional dynamic and the in-plane g-factor is anisotropic also for $(110)$-oriented QWs.


%

The effective g-factor tensor $\hat{g}^{*}$ in bulk GaAs is isotropic at the $\Gamma$-point 
but becomes increasingly anisotropic with the reduction of symmetry by heterostructure growth, potential gradients or low symmetry growth axis. The g-factor tensor reduces for asymmetric $(001)$ GaAs QWs and symmetric $(110)$ GaAs QWs to
\begin{equation}
{\small
\label{eq:gtensor}
\hat{g}^*_{C_{2v}^{001}}  = \left( {\begin{array}{*{20}c}
   {g_{s} } & {g_{a} } & 0  \\
   {g_{a} } & {g_{s} } & 0  \\
   0 & 0 & {g_{z} }  \\
\end{array}} \right),\,
\hat{g}^*_{C_{2v}^{110}} = \left( {\begin{array}{*{20}c}
   {g_{s} } & 0 & 0  \\
   0 &  {g_{s}  + 2g_a }& 0  \\
   0 & 0 & {g_{z} }  \\
\end{array}} \right),}
\end{equation}
where $g_{s}$ is the in-plane g-factor, $g_{a}$ the in-plane g-factor anisotropy, and $g_{z}$ the g-factor in growth direction. The \emph{in-plane} g-factor anisotropy based upon asymmetric $(001)$-oriented structures has been examined in detail in the past \cite{oestreichel1995, nefyodovprb2011, eldridgeprb2011} and a vast number of works exist on \emph{in/out-of-plane} g-factor anisotropy. 
However, symmetric, GaAs based $(110)$-oriented QWs have drawn a tremendous attention in the past due to vanishing Dresselhaus splitting for spins aligned along the growth direction \cite{dyakonovsps1986, ohnoprl1999, salisprl2001, dohrmannprl2004, glazovprb2010} and the g-factor tensor is defined according to Eq.~\ref{eq:gtensor} for $C_{2v}^{110} $ symmetry \footnote{The transformation into the [110]-coordinate system yields $x'=[001]$, $y'=[1\bar{1}0]$, $z'=[110]$.},i.e., $C_{2v}^{110} $ symmetry requires that only three independent diagonal entries of the g-factor tensor are non-zero.


In the following, we present detailed experimental measurements on the in-plane g-factor anisotropy in $(110)$-oriented QWs in dependence of the QW width and show that the high accuracy experiments are in excellent agreement with sophisticated $14\times 14 - k\cdot p$ calculations. The investigated sample is grown by molecular beam epitaxy and consists of ten undoped, symmetrical, $(110)$-oriented, GaAs/Al$_{0.32}$Ga$_{0.68}$As single QWs with thicknesses of 3, 4, 5, 6, 7, 8, 10, 12, 15, and 19 nm, respectively, separated by 80 nm barriers. The electronic wavefunction is subjected stronger to the 2D nature of the confining potential with decreasing well width, which in turn ultimatively reaches the value of the barriers and thus releases the wavefunction to three dimensionality again. From experiments measuring the exciton binding energy \cite{greenessc1983} this is expected to happen at a well width of about 4 nm for the given system.



\begin{figure}[tb]
  \centering
  \includegraphics[width=0.95 \columnwidth]{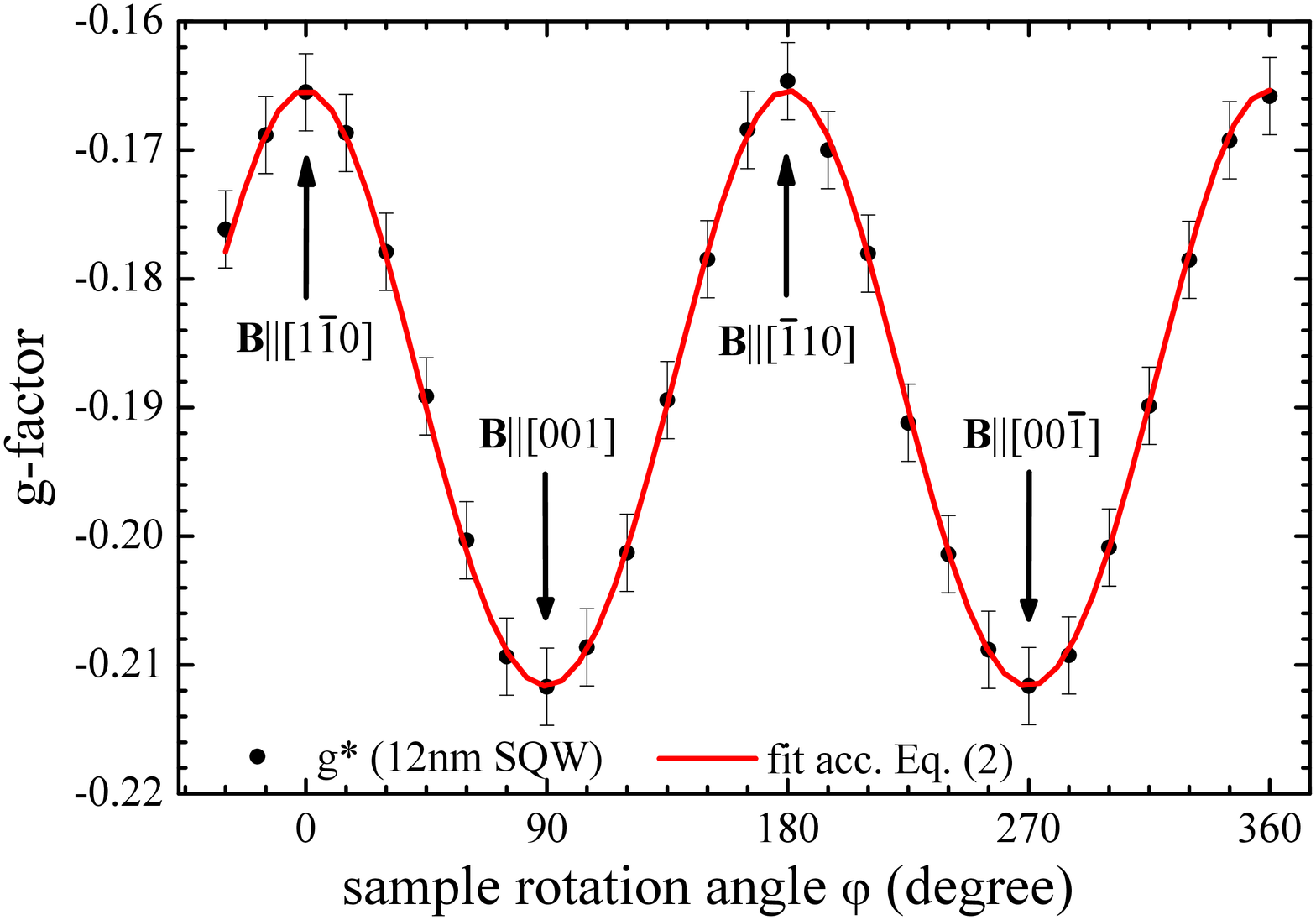}
  \caption{(Color online) Measured electron g-factor for the 12~nm, $(110)$-oriented, GaAs/Al${}_{0.32}$Ga${}_{0.68}$As, single QW in dependence on the angle $\varphi $ between the $[1\bar{1}0]$ in-plane axis and the in-plane magnetic field orientation (B=6T, T=20K) \cite{Note1}. 
The line is a fit to the data according to Eq.~\ref{eq:gmess} with $g_s = -0.212$, $g_a=+0.023 $ and $\phi = 6.96$~mrad.}
\label{fig:fig01}
\end{figure}

Information about the g-factor in GaAs QWs can by reliably accessed by measuring the polarization resolved time evolution of the photoluminescence of an optically excited spin polarization with an perpendicular magnetic field applied. This technique is known as spin quantum beat spectroscopy \cite{huebner2008}: The sample is mounted in Voigt geometry on a rotating sample holder with the growth and excitation axis perpendicular to the magnetic field axis in a helium flow cryostat with optical access within a split coil superconducting magnet. Spin polarized carriers are excited by circularly polarized laser pulses from an 80 MHz picosecond Ti:Sapphire laser and the photoluminescence is detected in backward direction and energy- and time resolved by a spectrometer and a synchroscan streak camera, respectively. A switchable retardation plate  and a polarizer perform the polarization resolution. Spin quantum beats occur due to the time evolution of the coherently excited Zeeman-split levels of spin-up and spin-down conduction band states. The beat (Larmor) frequency $\omega _{L} $ is directly linked to the electron g-factor $g^{*} $ and the magnetic field strength $B$ by $\omega _{L} =g^{*} \mu _{B} B\, /\hbar $. The hole spin dynamic is insignificant in the investigated experimental regime due to the fast hole spin relaxation times.


The measured effective g-factor $g^{*}$ is extracted from the polarization resolved intensity modulation for different orientations of the in-plane magnetic field. Figure~\ref{fig:fig01} shows the dependence of $g^{*}$ on the angle between the $[1\bar{1}0]$-axis and the in-plane magnetic field $\mathbf{B}=B_{0} (\sin \varphi ,\cos \varphi ,0)$.
The values for the symmetric $(g_{s})$ and antisymmetric $(g_{a} )$ contribution to $g^{*}$  are extracted according to the equation:
\begin{equation}\label{eq:gmess}
\begin{split}
g_{{\rm meas}}^*  & =   \pm \left| {\hat{g}_{C_{2v}^{110} }^*  \cdot \mathbf{B}} \right|/B_0 \\
&=  \pm \sqrt {g_s^{ 2}  + 2(g_s  + g_a )g_a \left( {1 +
{\cos(2}\varphi  + \phi_{0} )} \right)}.
\end{split}
\end{equation}
The angle $\phi_{0} $ is a free parameter which adjusts for the alignment mismatch of the sample with respect to the magnetic field axis.
\begin{figure}[tbp]
  \centering
  \includegraphics[width=0.95 \columnwidth]{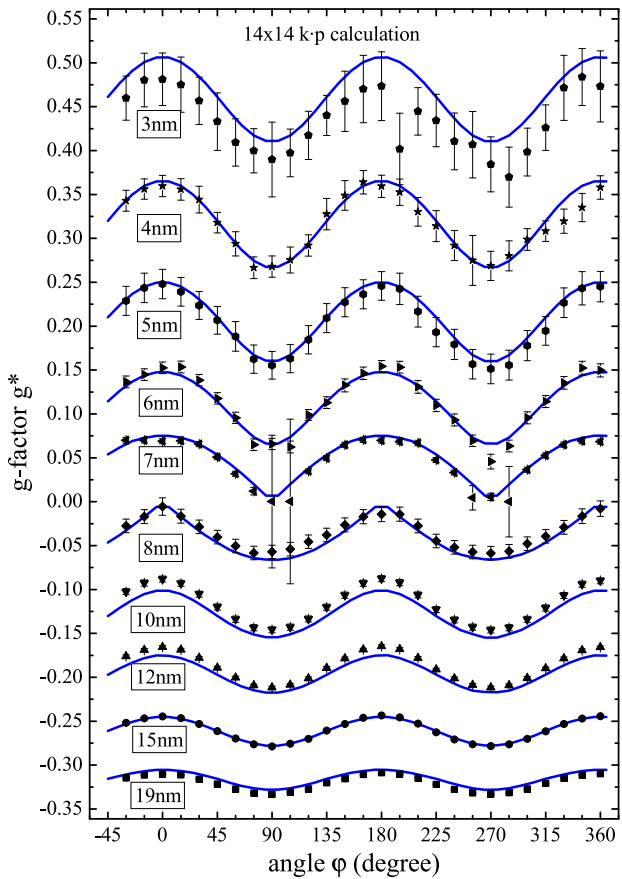}
  \caption{(Color online) Measured g-factors (squares) in dependence on the angle $\varphi $ for all ten QWs. The applied magnetic field is 6~T and the sample temperature 20~K. Note that the depicted values and curves are not shifted and all values correspond to the left axis. Calculations with a $14\times 14$ $k\cdot p$ model and a single common parameter set (solid lines).}
\label{fig:fig02}
\end{figure}

Figure~\ref{fig:fig02} shows the measured  g-factors (squares) for all ten QW widths measured simultaneous within the same sample in dependence on the angle $\varphi $. The measurement proves a significant in-plane g-factor anisotropy and shows a continuous increase of $g^{*}$ with decreasing QW width. With decreasing well width a) the Larmor precession frequency passes a minimum at a well width of about 7~nm b) the lifetime of the detected photoluminescence decreases due to higher electron-hole overlap and c) the spectra are stronger inhomogenously broadened due to growth imperfections. The inhomogeneous broadening affects the quality of the polarization resolved spin quantum beats and thus increases the error of the extracted g-factor. The values for $g_s$ and $g_a$ with decreasing QW width are extracted according to Eq.~\ref{eq:gmess} and depicted in Fig.~\ref{fig:fig03}. The sign of the measured g-factor depends on the energy-dependence of $g^{*}$ \cite{ivchenko1997, pfefferprb2006a, shichijjoap2009} and on the penetration of the wavefunction into the barrier material which has a positive g-factor. As a consequence $g_s$ increases monotonically with decreasing well width, i.e., increasing confinement energy. On the other hand $g_a$ reaches a maximum at a QW width of about 4~nm where the electronic wavefunction is most strongly localized in the quantization direction.


In the next section, we develop a theoretical description of the observed results based upon $k\cdot p$ perturbation theory. We follow the treatment of the fourteen band extended Kane model \cite{mayerprb1991, winkler2003} in which the spin-orbit interaction is included to calculate the dispersions and Zeeman splitting. Input parameters are the critical point energies, the interband matrix elements ($P,P',Q$), and k-linear terms due to SOI $(C_{k})$. The contributions from remote bands are included in the parameters $\gamma _{1,2,3}$ and $\kappa $ via $m^{*}$ and $g^{*}$ as described in \cite{mayerprb1991, winkler2003}. Magnetic interaction in the presence of an in-plane magnetic field $\mathbf{B}=(B_{x} ,B_{y} ,0)$ is taken into account by transformation of the quasi-momentum into the canonical momentum $\hat{k}_{x} = k_x+\frac{e}{\hbar}zB_y $ and $\hat{k}_{y} = k_y-\frac{e}{\hbar}zB_x $. We use the envelope function approximation for QW systems described by the effective-mass equation:
\begin{figure}[tbp]
  \centering
  \includegraphics[width=1.0 \columnwidth]{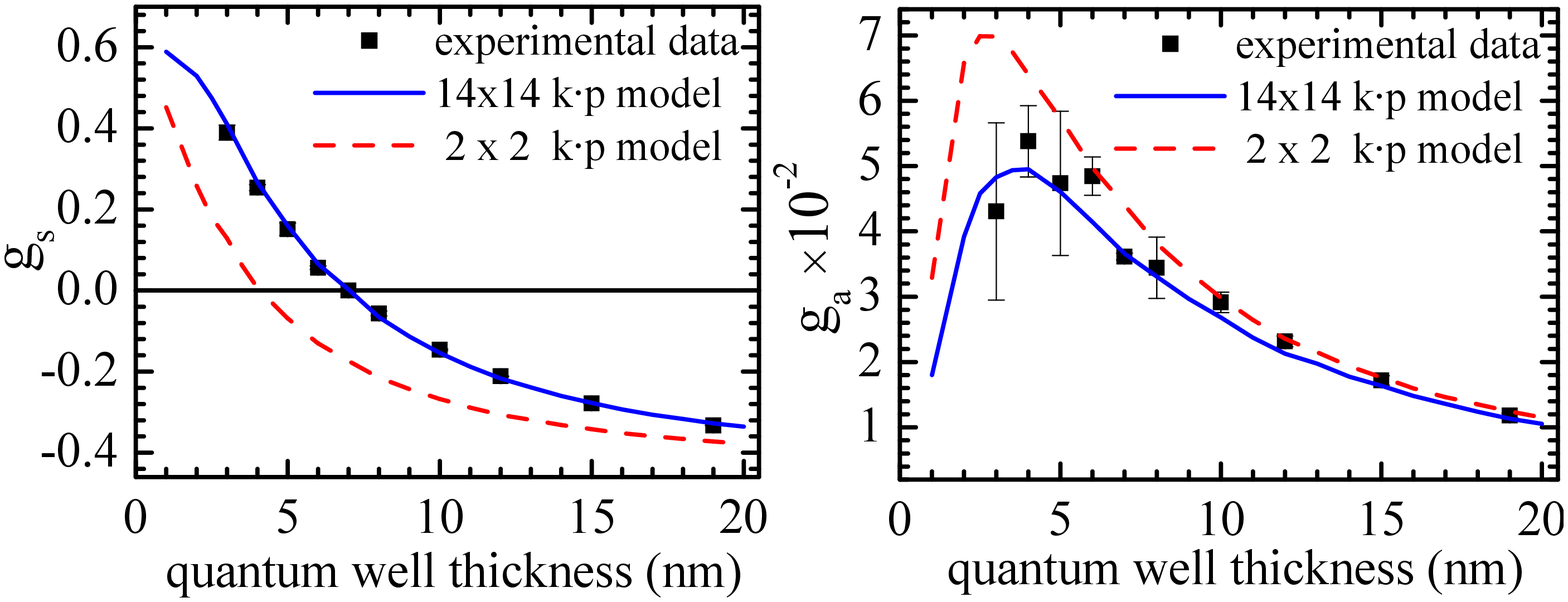}
  \caption{(Color online) The fitting values of $g_{s} $ (left) and $g_{a} $ (right) as a function of well width. The squares show the experimental data. Calculations by a $14\times 14$ Hamiltonian (solid lines), by a simplified $2\times 2$ Hamiltonian using fourth-order L{\"o}wdin perturbation theory (dashed lines).}
\label{fig:fig03}
\end{figure}
%
\begin{equation}\label{eq:Ham_eme}
\sum_{i=1}^{14}\left[H_{i,j}(\mathbf{\hat{k}})+V_i(z)\delta_{i,j}+H^Z_{i,j}\right]\psi_{i,{\mathbf
k}_\parallel}(z)=E\psi_{j,{\mathbf k}_\parallel}(z),
\end{equation}
where $H$ is the $k\cdot p$ Hamiltonian with $\hat{k}_{z} $ being replaced by the momentum operator $\hat{k}_z = -i\partial /\partial z$, $\mathbf{k}_{{\rm \parallel }} $ is the in-plane wave vector, $V(z)$ is the band offset potential, and $H^{Z} $ is an effective Zeeman Hamiltonian describing remote-band contributions.  Euler rotations of the coordinate system are applied to obtain the $14\times 14$ $k\cdot p$ Hamiltonian matrix for various crystallographic directions for the quantization axis ($z$-axis) \cite{ducprb2010}. We solve Eq.~\eqref{eq:Ham_eme} by expanding the envelope functions via a plane-wave basis \cite{allmenprb1992, elkurdiprb2003}. From the obtained band structure we compute the spin splitting between two spin states of the lowest conduction band $\Delta E_{k_{{\rm \parallel }} } =E_{k_{{\rm \parallel }} }^{(+)} -E_{k_{{\rm \parallel }} }^{(-)} $. This splitting includes for materials with bulk inversion asymmetry like GaAs both Dresselhaus and Zeeman splitting. At the band edge ($k_{\parallel } =0$), the Dresselhaus term vanishes and $\Delta E_{k_{{\rm \parallel }} =0} $ is a pure Zeeman splitting. The electron $g$-factor is extracted from the Zeeman splitting as $g^{*}=\Delta E_{k_{{\rm \parallel }} =0} /\mu _{B} B$.

Figure~\ref{fig:fig02} shows the calculated (solid lines) electron g-factor for all ten QWs as a function of the angle between magnetic field direction and $[1\bar{1}0]$-axis. The calculation is based on the full $14\times 14$ Hamiltonian using the band parameters listed in Table~\ref{table1}. The comparison shows an excellent agreement between theory and experiment.

Next, we carry out further analyses to understand the origin of the g-factor anisotropy within $k\cdot p$-theory. We use the L{\"o}wdin perturbation method \cite{winkler2003} to block-diagonalize the $14\times 14$ Hamiltonian and obtain a simplified $2\times 2$ Hamiltonian describing conduction band states. The terms $\hat{k}_{z}^{n} $ and $z^{n} $ are replaced by the expectation values $\langle \hat{k}_{z}^{n} \rangle $ and $\langle z^{n} \rangle $ of the quasi-two dimensional system. The expectation values with odd $n$ vanish in symmetric QWs, e.g., $\langle \hat{k}_{z} \rangle =0,$ $\langle z\rangle =0$. Up to fourth order of perturbation theory, the Hamiltonian for an electron in a QW is written as
$ H=H^{m^{*} } +H^{{\rm BIA}} +H^{B} $. The first term $H^{m^{*} } $ describes the parabolic dispersion with effective mass $m^{*} $. The second term $H^{{\rm BIA}} $ describes the Dresselhaus spin splitting due to the bulk inversion asymmetry. The third term $H^{B} $ represents the linear dependence of the Hamiltonian on magnetic field (terms of second and higher order in $B$ are neglected). The Hamiltonian $H^{B} $ takes for $k_{{\rm \parallel }} =0$ and the QW growth axis $z{\rm \parallel }[110]$ the form:
\begin{equation}\label{eq:B-dependence}
H^{B}=\left(\frac{g^*}{2}\mu_B+\alpha \langle
\hat{k}_z^2\rangle\right)\left(\sigma_x B_x+\sigma_y
B_y\right)+\beta\langle \hat{k}_z^2\rangle\sigma_y B_y,
\end{equation}
where $\sigma _{i} $ are the Pauli matrices. The term in Eq.~\eqref{eq:B-dependence} scaling with $\alpha$ describes the isotropic Zeeman splitting and the latter term in Eq.~\eqref{eq:B-dependence} describes the anisotropic Zeeman splitting scaling with:

{\small \begin{eqnarray}\label{eq:anisotropic}
&\beta &= \frac{e\,(3\hbar)^{-1}P^2Q^2}{E_g(E_g-E'_g-\Delta')}\left[\frac{4 E_g + \Delta}{E_g (E_g+\Delta)}-\frac{4 (E_g-E'_g-\Delta')}{E_g(E_g-E'_g)}\right]\nonumber\\
&+&
\frac{e\,(3\hbar)^{-1}P'^2Q^2}{(E_g-E'_g-\Delta')^{2}}\left[\frac{-3 E_g + \Delta}{E_g (E_g+\Delta)}+\frac{3 (E_g-E'_g-\Delta')}{(E_g-E'_g)E_g}\right].
\end{eqnarray}
}

We notice that for QWs grown with $z{\rm \parallel }[001]$, the Hamiltonian $H^{B} $ has the same isotropic term as in Eq.~\eqref{eq:B-dependence} but the anisotropic term vanishes, i.e., $\beta =0$. Furthermore, Eq.~\eqref{eq:anisotropic} perfectly demonstrates the link between the g-factor anisotropy and SOI since in the limit of zero spin-orbit gaps, $\Delta = \Delta' = 0$, all intricate g-factor peculiarities disappear leaving only the free electron g-factor.

\begin{table}[tbp]
\caption{Band parameters for GaAs and Al$_{0.32}$Ga$_{0.68}$As. $E_g, E'_g, \Delta, \Delta', \bar\Delta$ are in units of eV and $P, P', Q, C_k$ in eV$\cdot$nm and $m^{*}$ in $m_{0}$. The valence band offset is $\Delta E_{v}=0.35\Delta E_{g}$.}\label{table1}
\begin{tabular}{|c|cccccccc|}
\hline
      & $E_{g}$&$E'_{g}$&$\Delta$&$\Delta'$&$\bar{\Delta}$& $P$   & $P'$  & $Q$  \\ \hline
GaAs  & 1.517  & 4.504  & 0.341  & 0.171   & -0.05        & 1.049 & 0.445 & 0.821   \\ \hline
AlGaAs& 2.019  & 4.655  & 0.330  & 0.164   & -0.102       & 1.008 & 0.462 & 0.806  \\ \hline \hline
      & $-C_k$  & $m^{*}$& $\gamma _{1}$ &$\gamma_{2}$&$\gamma_{3}$ & $g^{*} $ & $\kappa $&\\ \hline
GaAs  & 0.00034 & 0.0665 & 6.98          &  2.06      & 2.93        & -0.44    & 1.2   & \\ \hline
AlGaAs& 0.00017 & 0.0927 & 5.95          &  1.66      & 2.45        & 0.60     & 0.54  &  \\ \hline
\end{tabular}
\end{table}

Diagonalizing the Hamiltonian in Eq.~\eqref{eq:B-dependence}, we obtain $\Delta E^{B}_{\mathbf{k}_\parallel=0} = g^{*}(\varphi )\mu_{B} B_0 $ where $g^{*}(\varphi ) $ matches the relation for $g^{*}_{\rm meas}$ in Eq.~\eqref{eq:gmess} with the analytical expressions for $g_s = {g^*}+\frac{2}{\mu_B}\alpha\langle \hat{k}_z^2 \rangle$ and $g_a = \frac{1}{\mu_B }\beta \langle \hat{k}_z^2\rangle$. The electron confined energy reduces to zero for well widths $d\to 0$ or $d\to \infty $, i.e., $\langle \hat{k}_{z}^{2} \rangle \to 0$, the anisotropic term vanishes, and the g-factor becomes isotropic again. The results obtained by the fourth order perturbation approach are depicted in Fig.~\ref{fig:fig03} and obviously higher order terms are necessary to correctly reproduce the symmetric $(g_s)$ and antisymmetric $(g_a)$ g-factor for the given parameter set.

We note that for diamond lattices (point group $O_{h}$) the terms $P',\bar{\Delta}$ and $C_{k}$ vanish \cite{winkler2003}. However $(110)$-grown heterostructures like Si/Ge/Si have the symmetry of the point group $D_{2h}$ and will still exhibit an anisotropic \emph{in-plane} g-factor as seen in the first term of Eq.~\eqref{eq:anisotropic} which are proportional to $P$ and $Q$ only. As a consequence the asymmetry in symmetric (110)-grown structures can be attributed to the interaction of the valence and upper conduction band states $(\propto Q)$, coupled to the lowest conduction band $(\propto P,P')$. We want to point out that theory also predicts a significant in-plane anisotropy of the effective mass of, e.g., 2.7\% for the 12~nm QW. An in-plane effective mass anisotropy has already been observed for \emph{asymmetric} (001)-grown QW structures \cite{rekerprl2002}.


In conclusion we investigated the anisotropy of the electron Land{\'e}-g-factor at low temperatures in symmetrically grown $(110)$-oriented GaAs/AlGaAs QWs via photoluminescence measurements. In contrast to asymmetric $(001)$-grown QWs with either a built-in potential gradient or an external applied electrical field the symmetry reduction inherently originates from the low symmetry growth direction of the QW structure. The g-factors for all QW widths are accurately modeled by $14\times 14$ $k\cdot p$ theory and the source terms for the in-plane anisotropy is extracted by fourth order perturbation theory.


We acknowledge financial support by the DFG within the priority program "SPP 1285 - Semiconductor Spintronics", the research group "Micro- and Nanostructures in Optoelectronics and Photonics" GRK 1464, ME 1916/2, FO 637/1, and the excellence cluster "QUEST".

\bibliography{110-gxy-Paper}

\end{document}